\documentclass[prl,aps,showpacs,epsf, twocolumn]{revtex4}
\usepackage{graphicx}

\begin{document}
\title{Resonant scattering properties close to a p-wave Feshbach resonance}
\author{F.\,Chevy$^1$, E.\,G.\,M.\,
van Kempen$^2$, T.\,Bourdel$^1$, J. Zhang$^{3}$,
L.\,Khaykovich$^{4}$,  M.\,Teichmann$^1$, L.\,Tarruell$^1$,
S.\,J.\,J.\,M.\,F.\,Kokkelmans$^{2}$,  and C.\,Salomon$^1$}
\affiliation{$^1$ Laboratoire Kastler-Brossel, ENS, 24 rue
Lhomond, 75005 Paris} \affiliation{$^2$ Eindhoven University of
Technology, P.O.~Box~513, 5600~MB Eindhoven, The
Netherlands}\affiliation{$^3$ SKLQOQOD, Institute of
Opto-Electronics, Shanxi University, Taiyuan 030006, P.R. China}
\affiliation{$^4$ Department of Physics, Bar Ilan University,
Ramat Gan 52900, Israel.}\date{\today}

\begin{abstract}
We present a semi-analytical treatment of both the elastic and
inelastic collisional properties near a p-wave Feshbach resonance.
Our model is based on a simple three channel system that
reproduces  more elaborate coupled-channel calculations. We stress
the main differences between s-wave and p-wave scattering. We show
in particular that, for elastic and inelastic scattering close to
a p-wave Feshbach resonance, resonant processes dominate over the
low-energy behaviour.

\end{abstract}


\maketitle

\section{Introduction}

The observation of molecular gaseous Bose-Einstein condensates
(BEC) and the subsequent experimental study of the BEC-BCS
crossover \cite{Jochim03,Zwirlein03,Bourdel04,Kinast04,Greiner03}
were made possible by the possibility of tuning interatomic
interactions using a magnetic field (the so-called Feshbach
resonances). Although all these experiments were  based on s-wave
interatomic interactions, it is known from condensed matter
physics that
 superfluidity of fermionic systems can also arise through
higher order partial waves. The most famous examples of this
non-conventionnal superfluidity are  $^3$He \cite{Lee97}, for
which the Cooper pairs spawn from p-wave interactions, and
high-$T_{\rm c}$ superconductivity, in which pairs are known to
possess  d-wave symmetry \cite{Tsuei00}. Recent interest in p-wave
interactions in cold atom gases stemmed from these possibilities
and resulted in the observation of p-wave Feshbach resonances in
$^{40}$K \cite{Regal03} and $^6$Li \cite{Zhang04,Ketterle}, as
well as
 theoretical studies on the superfluidity of cold atom
interacting through p-wave pairing \cite{Ho04,Ohashi04}.

The present paper is devoted to the study of p-wave interactions
close to a Feshbach resonance and it  derives some results
presented in \cite{Zhang04}. In a first part, we present the model
we use to describe both elastic and inelastic processes that are
discussed in the second part. We stress in particular the main
qualitative differences between p-wave and s-wave physics and show
 that contrarily to the case of s-wave that is
dominated by low energy physics,  p-wave scattering is dominated
by a resonance peak associated to the quasi bound molecular state.
Finally, we compare our analytical results to numerical
coupled-channel calculations.

\section{Model for p-wave interactions}

We consider the scattering of two identical  particles of mass
$m$. As usual when treating a two-body problem, we work in the
center of mass frame and only consider the motion of a fictitious
particle of mass $m/2$ interacting with a static potential. In
order to study the p-wave Feshbach resonance, we use a model based
on the separation of  open and closed channels. In this framework,
the Feshbach resonance arises in an open channel as a result of
the coupling with a closed channel \cite{Taylor}. At resonance,
scattering properties are dominated by resonant effects and we can
neglect all ``background" scattering (ie. we assume there is no
scattering far from resonance).

\begin{figure}
\centerline{\includegraphics{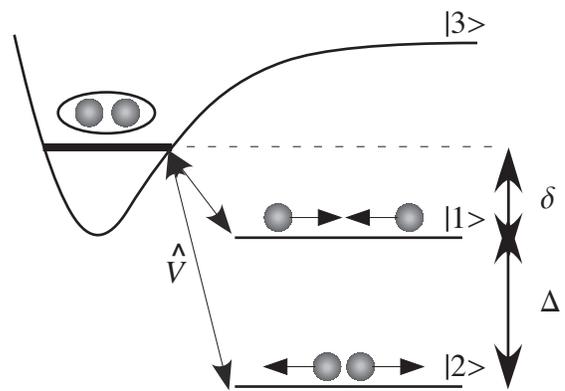}} \caption{The p-wave model:
we consider three internal states, labelled $|1\rangle$,
$|2\rangle$ and $|3\rangle$. States $|1\rangle$ and $|2\rangle$
are two open channels corresponding respectively to the incoming
and decay channels. The released energy in an inelastic collision
bringing atom initially in $|1\rangle$ to $|2\rangle$ is denoted
$\Delta$. State $|3\rangle$ is a closed channel that possesses a
p-wave bound state of energy $\delta$ nearly resonant with state
$|1\rangle$. Finally, we assume that these three channels interact
through a potential $\widehat V$ acting only on the internal
states and coupling the two open channels to the closed channels.}
\label{FigScheme}
\end{figure}

\begin{enumerate}
\item We restrict ourselves to a 3-channel system, labelled 1,2
and 3 which correspond to the different two-body spin
configurations (Fig. \ref{FigScheme}). Channels $|1\rangle$ and
$|2\rangle$ are open channels. We focus on the situation were
atoms are prepared initially in state $|1\rangle$. Atoms may be
transferred to state $|2\rangle$ after an inelastic process.
Channel $|3\rangle$ is closed and hosts the bound state leading to
the Feshbach resonance.

Let us consider for instance the case of $^6$Li atoms prepared in
a mixture of $|F=1/2,m_F=1/2\rangle$ and
$|F'=1/2,m'_F=-1/2\rangle$. In this system, the only two-body
decay channel is associated with the flipping of a $m'_F=-1/2$
atom to $m'_F=1/2$. If we denote by $(m_F,m'_F)$ the symmetrized
linear combination of the states $|F=1/2,m_F\rangle$ and
$|F'=1/2,m'_F\rangle$, then $|1\rangle=(1/2,-1/2)$ and
$|2\rangle=(1/2,1/2)$.

\item \label{Ass2} The Feshbach resonances studied here are all
located at values of the magnetic field where the Zeeman splitting
is much larger than the hyperfine structure. In first
approximation we can therefore assume that the internal states are
 described by uncoupled electronic and nuclear spin states. In the absence of any dipolar or hyperfine coupling
 between the electronic singlet and triplet manifolds,
 we assume we have no direct interaction in channels 1 and 2 so that the eigenstates are
plane waves characterized by their relative wave-vector $k$ and
their energy $E_1(k)=\hbar^2 k^2/m$ (channel 1) and
$E_2(k)=-\Delta+\hbar^2 k^2/m$ (channel 2). $\Delta>0$ is the
energy released in an inelastic process leading from 1 to 2.
$\Delta$ can be considered  as independent  of the magnetic field
and is assumed to be much larger than any other energy scales (in
the  case relevant to our experiments, $\Delta/h\sim 80\,$MHz is
the hyperfine splitting of $^6$Li at high field).

\item In channel 3, we consider only a p-wave bound state nesting
at an energy $\delta$ quasi-resonant with channel 1. In the case
of $^6$Li atoms in the $F=1/2$ hyperfine state, $\delta=2\mu_B
(B-B_0)$, where $B$ is the magnetic field and $B_0$ is the
position of the ``bare" Feshbach resonance. If  the projection of
the angular momentum (in unit of $\hbar$)  is denoted by $m_{\bf
u}$ for a quantization axis chosen along some vector $\bf u$, the
eigenfunctions associated to this bound state can be written as
$g(r) Y_1^{m_{\bf u}}(\theta,\phi)$, where $(r,\theta,\phi)$ is
the set of polar coordinates and the $Y_l^m$ are the spherical
harmonics.

\item \label{Ass3}The coupling $\widehat V$ between the various
channels affects only the spin degrees of freedom. Therefore the
orbital angular momentum is conserved during the scattering
process and we restrict our analysis to the p-wave manifold. This
is incontrast to the situation in heavy alkalis where incoming
particles in s-wave can be coupled to molecular states of higher
orbital angular momentum \cite{Leo00, Voltz04}.

We assume also that the only non-vanishing matrix elements are
between the closed and the open channels (ie. $\langle
1,2|\widehat V|3\rangle$ and $\langle 3|\widehat V|1,2\rangle$).

\end{enumerate}

Let us denote a state of the system by $|\alpha,\chi\rangle$,
where $\alpha\in \{1,2,3\}$ and $\chi$ describe respectively the
internal (spin) and the orbital degrees of freedom. According to
assumption (\ref{Ass3}), the matrix element $\langle
\alpha,\chi|\widehat V|\alpha',\chi'\rangle$ is simply given by:

\begin{equation}
\langle \alpha,\chi|\widehat
V|\alpha',\chi'\rangle=\langle\chi|\chi'\rangle \langle
\alpha|\widehat V|\alpha'\rangle, \label{EqnOverlap}
\end{equation}

\noindent and is therefore simply proportional to the overlap
$\langle\chi|\chi'\rangle$ between the external states.

Let us now particularize to the case where $\alpha\in \{1,2\}$,
and $|\chi\rangle=|{\bf k}\rangle$ is associated to a plane wave
of relative  momentum $\hbar {\bf k}$. According to hypothesis
(\ref{Ass3}), this state is coupled only to the closed channel
$|3\rangle$. Moreover, using the well known formula
$e^{ikz}=\sum_l i^l \sqrt{4\pi (2 l+1)} j_l (kr) Y_l^{0}
(\theta,\phi)$, where the $j_l$ are the spherical Bessel
functions, we see that $|\chi\rangle$ is coupled only to the state
 $|\alpha'=3,\rm k=0\rangle$ describing the pair in the
bound state $|\alpha'=3\rangle$ with zero angular momentum in the
$\bf k$ direction. The matrix element then reads

\begin{equation}
\langle \alpha,{\bf k}|\widehat V|3,m_{\bf
k}\rangle=i\delta_{m_{\bf k},0} \sqrt{\frac{12\pi}{L^3}}
\langle\alpha|\widehat V|3\rangle\int g^*(r)j_1 (kr)r^2\,{\rm d}r,
\label{EqnF}
\end{equation}

\noindent where $L^3$ is a quantization volume. Since for small
$k$ we have $j_1 (kr)\sim kr/3$, the matrix element $\langle
\alpha,{\bf k}|\widehat V|3,m_{\bf k}\rangle$ takes the general
form

\begin{equation}\langle \alpha,{\bf k}|\widehat V|3,m_{\bf
k}\rangle=\delta_{m_{\bf k},0} \frac{k
F_\alpha(k)}{\sqrt{L^3}},\end{equation}

\noindent where $F_\alpha (k)$ has a finite (in general non zero)
limit when $k$ goes to zero.

Later on, we shall also need the coupling between  $|\alpha,{\bf
k}\rangle$ and $|3,m_{\bf k'}=0\rangle$ (that will be denoted by
$|3,0_{\bf k'}\rangle$), where the momentum $\bf k$ and the
direction of quantization $\bf k'$ are no longer parallel. The
calculation presented above yields readily

\begin{equation}
\langle \alpha,{\bf k}|\widehat V|3,0_{\bf k'}\rangle=\frac{k
F_\alpha(k)}{\sqrt{L^3}} \langle 0_{\bf k}|0_{\bf
k'}\rangle=\frac{k F_\alpha(k)}{\sqrt{L^3}}\cos (\widehat{{\bf
k},{\bf k'}}),
\end{equation}

\noindent where $(\widehat{{\bf k},{\bf k'}})$ is the angle
between ${\bf k}$ and ${\bf k}'$ \cite{Edmunds}.

\begin{figure*}
\scalebox{.75}{\includegraphics{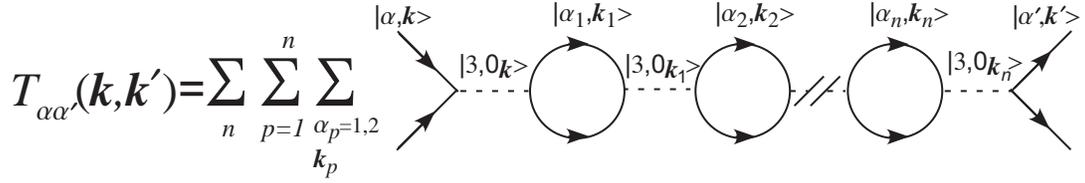}} \caption{Diagrammatic
expansion of the $T$-matrix. The full lines  (resp. dashed)
represent free atoms (resp. molecules). $|\alpha, \bf k \rangle$
is the scattering state of the two particles, in the internal
state $\alpha=1,2$. $|3, 0_{\bf k }\rangle$ represents the state
of a p-wave molecule with  orbital angular momentum component zero
on ${\bf k}$ direction. }\label{FitTSum}
\end{figure*}

\section{T-Matrix}

From general quantum theory, it is known that the scattering
properties of a system are given by the so-called T-matrix
$\widehat T$. It can be shown in particular that $\widehat T$  is
given by the following  expansion in power of the coupling
potential

\begin{equation}\widehat T (E)=\widehat V+\widehat V\widehat G_0 (E)\widehat
V+\widehat V\widehat G_0 (E)\widehat V\widehat G_0 (E)\widehat V
+..., \label{EqnExpansion}
\end{equation}

\noindent where $\widehat G_0 (E)=1/(E-\widehat H_0)$ and
$\widehat H_0=\widehat H-\widehat V$ is the free hamiltonian of
the system.

Let us consider $|\alpha, {\bf k}\rangle$ and $|\alpha',{\bf
k}'\rangle$, two states of the {\em open channels} and let us set
$T_{\alpha\alpha'}({\bf k},{\bf k}',E)=\langle\alpha,{\bf
k}|\widehat T (E)|\alpha',{\bf k}'\rangle$. According to formula
(\ref{EqnExpansion}), this matrix element is the sum of terms that
can be represented by the diagram of Fig. \ref{FitTSum} and we get
after a straightforward calculation

$$T_{\alpha\alpha'}({\bf k},{\bf
k}',E)=\frac{kk'}{L^3}F_\alpha (k)F_{\alpha'}^* (
k')\sum_{n=0}^\infty R_n\Lambda (E)^n G_0^{(m)} (E)^{n+1}.$$

Here $G_0^{(m)} (E)=1/(E-\delta)$ is the free propagator for the
molecule, $\Lambda=\Lambda_1+\Lambda_2$ with

\begin{equation}\Lambda_\alpha (E)=\int{\frac{q^4 dq}{(2\pi)^3}
\frac{|F_\alpha (q)|^2}{E-E_\alpha (q)}} \label{EqnLambda}
\end{equation}

\noindent results of the integration on the loops, and finally

$$R_n=\int d^2\Omega_1...d^2\Omega_n \cos (\widehat{{\bf
k},{\bf k}_1})\cos (\widehat{{\bf k}_1,{\bf k}_2})...\cos
(\widehat{{\bf k}_n,{\bf k'}}),$$

\noindent where $\Omega_p$ is the solid angle associated to ${\bf
k}_p$, arises from the pair breaking vertices $|3,0_{{\bf
k}_i}\rangle\rightarrow |\alpha_{i+1},{\bf k}_{i+1}\rangle$. This
last integral can be calculated recursively and we get
$R_n=(4\pi/3)^n\cos (\widehat{{\bf k},{\bf k}'})$, that is for the
$T$-matrix

$$T_{\alpha\alpha'}({\bf k},{\bf
k}',E)=\frac{1}{L^3}\frac{kk'F_\alpha (k)F_{\alpha'}^* (
k')}{E-\delta-\Sigma_1-\Sigma_2}\cos(\widehat{{\bf k},{\bf
k}'}),$$

\noindent with $\Sigma_\alpha=4\pi\Lambda_\alpha/3$.

This expression can be further simplified since, according to Eqn.
(\ref{EqnF}), the width of $F_\alpha (q)$ is of the order of
$1/R_e$, where $R_e$ is the characteristic size of the resonant
bound state. In the low temperature limit, we can therefore expand
$\Sigma_\alpha$ with the small parameter $kR_e$.

From Eqn. (\ref{EqnF}), we see that replacing $F_\alpha (q)$ by
its value at $q=0$ leads to a $q^2$ divergence. This divergence
can be regularized by the use of counter terms in the integral,
namely by writing that

$$
\begin{array}{l}
\Sigma_\alpha (E)=\\
\displaystyle{\int{ |F_\alpha (q)|^2\left [\frac{q^4}{E-E_\alpha
(q)}+\frac{mq^2}{\hbar^2}+\frac{m^2}{\hbar^4}(E-E_\alpha
(0))\right ] \frac{dq}{6\pi^2}}}\\
\displaystyle{-\int{|F_\alpha
(q)|^2\frac{mq^2}{\hbar^2}\frac{dq}{6\pi^2}}-(E-E_\alpha
(0))\int{|F_\alpha (q)|^2\frac{m^2}{\hbar^4}\frac{dq}{6\pi^2}}},
\end{array}$$

\noindent where we have assumed that $F_\alpha$ was decreasing
fast enough at large $q$ to ensure the convergence of the
integrals. $|F_\alpha (q)|^2$ can now be safely replaced by
$\lambda_\alpha=|F_\alpha (0)|^2$ in the first integral and we
finally get

$$\Sigma_\alpha=-i\frac{\lambda_\alpha}{6\pi}\frac{m}{\hbar^2}\left(\frac{m}{\hbar^2}(E-E_\alpha
(0))\right)^{3/2}-\delta_{0,\alpha}-\eta_\alpha (E-E_\alpha
(0)),$$

\noindent with

\begin{eqnarray*}
\delta_0^{(\alpha)}&=&\int{|F_\alpha (q)|^2\frac{mq^2}{\hbar^2}\frac{dq}{6\pi^2}}\\
\eta_\alpha&=&\int{|F_\alpha
(q)|^2\frac{m^2}{\hbar^4}\frac{dq}{6\pi^2}}.
\end{eqnarray*}

If we assume that the release energy $\Delta$ is much larger than
$E$ and if we set $\delta_0=\delta_0^{(1)}+\delta_0^{(2)}$ and
$\eta=\eta_1+\eta_2$, we get for the $T$-matrix

\begin{equation}T_{\alpha\alpha'}({\bf k},{\bf
k}',E)\simeq\frac{1}{L^3}\frac{kk'F_\alpha (0)F_{\alpha'}^*
(0)\cos(\widehat{{\bf k},{\bf
k}'})/(1+\eta)}{E-\tilde\delta+i\hbar\gamma (E)/2}. \label{EqnT}
\end{equation}

with

\begin{eqnarray*}
 \hbar\gamma(E)&=&\left(\frac{m}{
\hbar^2}\right)^{5/2}\frac{\left(\lambda_2\Delta^{3/2}+\lambda_1
E^{3/2}\right)}{3\pi (1+\eta)}\\
\tilde\delta&=&\frac{(\delta-\delta_0)}{1+\eta}.
\end{eqnarray*}

We note that this expression for the T-matrix is consistent with
the general theory of multichannel scattering resonances
\cite{Taylor} , where resonantly enhanced transitions to other
channels are readily included. In a similar context of two open
channels and a Feshbach resonance, a recent experiment was
analyzed \cite{Voltz04}, that involved the decay of a molecular
state formed from a Bose-Einstein condensate.

\section{S-wave vs p-wave}

This section is devoted to the discussion of the expression found
for the T-matrix. In addition to the scattering cross section, the
study of the T-matrix yields important informations on the
structure of the dressed molecular state underlying the Feshbach
resonance and we will demonstrate important qualitative
differences between the behaviours of p-wave and s-wave
resonances.

{\bf Molecular state}. The binding energy $E_b$ of the molecule is
given by the pole of $T$. In the limit $\delta\sim\delta_0$, it is
therefore given by:

$$E_{\rm
b}=\tilde\delta-i\hbar\gamma (\tilde\delta)/2,$$

 We see
that the real part of $E_b$ (the ``physical" binding energy) is
 $\sim\tilde\delta$ and therefore scales linearly with the
detuning $\delta-\delta_0$. This scaling is very different from
what happens for s-wave processes where we expect a
$(\delta-\delta_0)^2$ behavior. This difference is in practice
very important: indeed, the molecules can be trapped after their
formation only if their binding energy is smaller than the trap
depth. The scaling we get for the p-wave molecules means that the
binding energy increases much faster when we increase the detuning
than what we obtain for s-wave molecules (this feature was already
pointed out in \cite{Ho04}). Hence, p-wave molecules must be
looked for only in the close vicinity of the Feshbach resonance --
for instance, for $\eta\ll 1$ (relevant for $^6$Li, as we show
below) and a trap depth of 100~$\mu$K, the maximum detuning at
which molecules can be trapped is $\simeq 0.1$~G.

This asymptotic behavior of the binding energy is closely related
to the internal structure of the molecule. Indeed, the molecular
wave function $|\psi_{\rm m} (B)\rangle$ can be written as a sum
$|{\rm open}\rangle+|{\rm closed}\rangle$ of its projections on
the closed and open channels, that correspond respectively to
short and long range molecular states. If we neglect decay
processes by setting $\lambda_2=0$, we can define the magnetic
moment of the molecule (relative to that of the free atom pair) by

$$\Delta\mu_{\rm eff} (B)=-\frac{\partial E_{\rm b}}{\partial B}=-\frac{\partial \tilde\delta}{\partial
B},$$

\noindent that is, in the case of $^6$Li where $\delta=2\mu_B
(B-B_0)$,

\begin{equation}
\Delta\mu_{\rm eff} (B)=-\frac{2\mu_B}{1+\eta}. \label{EqnMag1}
\end{equation}

On the other hand, we can also write $E_b=\langle \psi_{\rm m}
(B)|\widehat H (B)|\psi_{\rm m}(B)\rangle$. Since in the absence
of any decay channel, the molecular state is the ground state of
the two-body system, we can write using the Hellman-Feynman
relation

$$\Delta\mu_{\rm eff}=-\frac{\partial E_b}{\partial B}=-\langle \psi_{\rm m} (B)|\frac{\partial\widehat H (B)}{\partial B}|\psi_{\rm
m}(B)\rangle.$$

In our model, the only term of the hamiltonian
 depending on the magnetic field is the energy $\delta=2\mu_{\rm B} (B-B_0)$ of the
 bare molecular state in the closed channel and we finally have

\begin{equation}
\Delta\mu_{\rm eff}=-2\mu_B\langle{\rm closed}|{\rm
closed}\rangle. \label{EqnMag2}
\end{equation}

If we compare Eqn. (\ref{EqnMag1}) and (\ref{EqnMag2}), we see
that the probability $P_{\rm closed}=\langle{\rm closed}|{\rm
closed}\rangle$ to be in the closed channel is given by

$$P_{\rm closed}=1/(1+\eta).$$

In other words, unless $\eta=\infty$, there is always a finite
fraction of the wave-function in the tightly bound state. In
practice, we will see that in the case of $^6$Li, $\eta\ll 1$.
This means that the molecular states that are nucleated close to a
Feshbach resonance are essentially short range molecules. On the
contrary, for s-wave molecules, $E_{\rm b} \propto
(\delta-\delta_0)^2$ leads to $\Delta\mu_{\rm
eff}=-2\mu_B\langle{\rm closed}|{\rm closed}\rangle \propto
(\delta-\delta_0)$. This scaling leads to a zero probability of
occupying the bare molecular state near a s-wave resonance.

\begin{figure}
\includegraphics[width=\columnwidth]{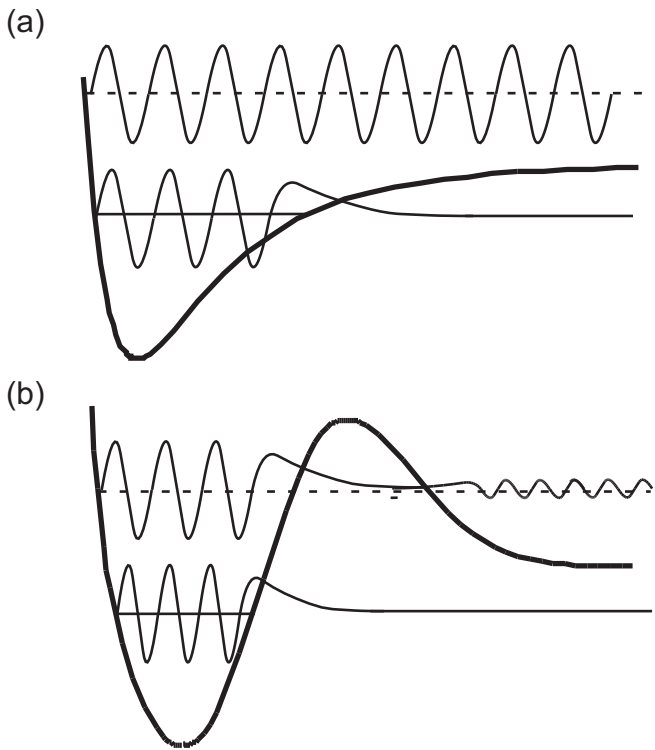}
\caption{Effect of the centrifugal barrier on the bound state in
p-wave Feshbach resonances. (a) Case of a s-wave scattering: the
bare molecular state goes from $\delta<0$ (full line) to
$\delta>0$ (dashed line). In the process, the molecular state
becomes unstable and the wave function becomes unbounded. (b) In
the case of p-wave bound state, the presence of the centrifugal
barrier smoothes the transition from $\delta>0$ to $\delta<0$.
Even for $\delta>0$, the wave-function stays located close to the
bottom of the well.} \label{FigCentrifugal}
\end{figure}

We can illustrate this different behaviours  in the simplified
picture of Fig. \ref{FigCentrifugal}. For small detunings around
threshold, the p-wave potential barrier provides a  large
forbidden region, that confines the bound state behind this
barrier. The bound state wavefunction decays exponentially inside
the barrier and the tunneling remains nearly negligible. Since
there is no significant difference for the shape of the p-wave
bound state for small positive and negative detunings, the linear
dependence of the closed channel with magnetic field will be
conserved for the bound state, and therefore the binding energy
will also linearly approach the threshold. We note that the linear
dependence close to threshold can also be found from the general
Breit-Wigner expression for a resonance, in combination with the
p-wave threshold behavior of the phase shift \cite{Taylor}.

The imaginary part of $E_b$ corresponds to the lifetime of the
molecule. In the case of s-wave molecules for which two-body decay
is forbidden \cite{Stoof88,Dieckmann02}, the only source of
instability is the coupling with the continuum of the incoming
channel that leads to a spontaneous decay when $\tilde\delta>0$
(see Fig. \ref{FigCentrifugal}.a). By contrast, we get a finite
lifetime in p-wave even at $\tilde\delta<0$: due to dipolar
relaxation between its constituents, the molecule can indeed
spontaneously decay towards state $|2\rangle$. For
$\tilde\delta\sim 0$, the decay rate $\gamma_0$ associated to this
process is given by:

$$\gamma_0=\gamma(0)=\frac{\lambda_2}{1+\eta}\frac{m}{3\pi\hbar^3}\left(\frac{m\Delta}{
\hbar^2}\right)^{3/2}.$$

\noindent {\bf Elastic scattering}. The scattering amplitude
$f({\bf k},{\bf k}')$ for atoms colliding in the channel 1 can be
extracted from the $T$-matrix using the relation

$$f({\bf k},{\bf k}')=-\frac{m L^3}{4\pi\hbar^2}T_{11}({\bf k},{\bf
k}',E=\hbar^2k^2/m),$$

\noindent that is

$$f({\bf k},{\bf k}')=-\frac{m\lambda_1}{4\pi\hbar^2}\left(\frac{k^2\cos(\widehat{{\bf k},{\bf k}'})/(1+\eta)}{
\hbar^2 k^2/m-\tilde\delta-i\hbar\gamma/2}\right).$$

The $\cos(\widehat{{\bf k},{\bf k}'})$ dependence is
characteristic of p-wave processes and, once again, this
expression shows dramatic differences with the s-wave behavior.
First, at low $k$, $f({\bf k},{\bf k}')$ vanishes like $k^2$. If
we introduce the so-called scattering volume $V_s$ \cite{Taylor}
defined by

$$f({\bf k},{\bf k}')=-V_s k^2,$$

\noindent then we have

$$V_s=\frac{-m\lambda_1}{4\pi\hbar^2 (\delta-\delta_0+i(1+\eta)\hbar\gamma_0/2)}\sim \frac{-m\lambda_1}{2\pi\hbar^2 (\delta-\delta_0)},$$

\noindent if we neglect the spontaneous decay of the molecule. We
see that in this approximation, the binding energy $E_b$ of the
molecule is given by

$$E_{\rm b}=-\frac{m\lambda_1}{2\pi \hbar^2 (1+\eta)}\frac{1}{V_s}.$$

In s-wave processes, the binding energy and the scattering length
are related  through the universal formula $E_{\rm
b}=-\hbar^2/ma^2$. This relationship is of great importance since
it allows to describe both scattering properties and the molecular
state with  the sole  scattering length, without having to care
with any other  detail of the interatomic potential. In the case
of p-wave, we see that no such universal relation exists between
the scattering volume and $E_{\rm b}$, a consequence of the fact
that we have to deal with short range molecular states, even at
resonance. We therefore need two independent parameters to
describe both the bound states and the scattering properties.

 In the general case, the
elastic cross-section $\sigma_{\rm el}$ is proportional to
$|f|^2$. According to our calculation, we can put $\sigma_{\rm
el}$ under the general form

\begin{equation}
\sigma_{\rm
el}(E)=\frac{CE^2}{(E-\tilde\delta)^2+\hbar^2\gamma^2/4},
\label{EqnCrossSection}
\end{equation}

\noindent where $E=\hbar^2 k^2/m$ is the kinetic energy of the
relative motion  and $C$ is a constant depending on the
microscopic details of the system. Noticeably, the energy
dependence of the cross-section exhibits a resonant behavior at
$E=\tilde\delta$ as well as a plateau when $E\rightarrow \infty$,
two features that were observed in the numerical coupled channel
calculations presented in \cite{Ticknor04}. Once again, this leads
to physical processes very different from what is expected in
s-wave scattering. Indeed, we know that in s-wave, we have $f\sim
-a$, as long as $ka\ll 1$. Since $a$ is in general non zero, the
low energy behavior gives a non negligible contribution to the
scattering processes. By contrast, we have just seen that in the
case of p-wave, the low energy contribution was vanishingly small
($\sigma_{\rm el}\sim E^2$) so that the scattering will be
dominated by the resonant peak $E\sim\tilde\delta$.

\noindent {\bf Inelastic scattering}. For two particules colliding
in channel 1 with an energy $E=\hbar^2k^2/m$, the probability to
decay to channel 2 is proportional to $\rho_2 (k') |T_{12}({\bf k
},{\bf k}',E)|^2$ where $\rho_2$ is the density of state in
channel 2. Since $k'$ is given by the energy conservation
condition $\hbar^2 k'^2/m-\Delta=E$, and in practice $\Delta \gg
E$, we see that $k'\sim \sqrt{m \Delta/\hbar^2}$ is therefore a
constant. Using this approximation, we can write the 2-body loss
rate $g_2 (E)$ for particles of energy $E$ as:

$$g_2 (E)=\frac{D E}{(E-\tilde\delta)^2+\hbar^2\gamma^2/4},$$

\noindent where $D$ is a constant encapsulating the microscopic
details of the potential.

\begin{table*}
\begin{ruledtabular}
\centerline{\begin{tabular}{c|ccc|ccc}
channel&&(1/2,-1/2)&&&(-1/2,-1/2)&\\
\hline
$m_l$&-1&0&1&-1&0&1\\
\hline $C\,(10^{-13}\,{\rm cm}^2)$
&0.22&0.22&0.22&0.87&0.88&0.87\\
$D\,(10^{-13}\,{\rm cm}^2\mu{\rm K}/{\rm s})$&0.00002&0.59&0.56&$3\times 10^{-5}$&1.54&5.72\\
$\gamma_{0}\,({\rm s}^{-1})$&$<10^{-2}$&$110$&$110$&$<1$&220&830\\
$a\,(\mu {\rm
K}^{-1/2})$&0.0017&0.0017&0.0017&0.0024&0.0024&0.0024\\
$\eta$ &0.22&0.22&0.22&0.23&0.23&0.23\\
$\delta B_{\rm F}$\,(G)&-0.0036&0&-0.0036&-0.012&0&-0.012
\end{tabular}}
\caption{\label{table1} Values of the phenomenological parameters
obtained after a fit to the coupled channel calculations data of
Fig. \ref{FigClosedChannel}. $\delta B_{\rm F}$ is the shift
between the $m_l=\pm 1$ and $m_l=0$ resonances.}
\end{ruledtabular}
\end{table*}

\section{Comparison with coupled-channel calculation}

\begin{figure}
\includegraphics[width=\columnwidth]{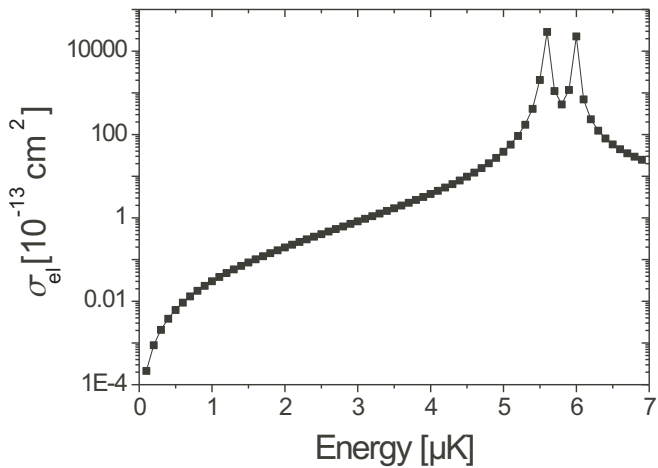} \caption{Energy dependence of the elastic cross section.
Dots: numerical closed channel calculation. The left peak
corresponds to  $m_l=0$ and the right peak to $m_l=\pm 1$. Full
line: Fits using Eqn. (\ref{EqnCrossSection2}) }
\label{FigClosedChannel}
\end{figure}

The quantities such as  $C$, $D$, $\gamma_0$ etc. that were
introduced in the previous section were only phenomenological
parameters to which we need to attribute some value to be able to
perform any comparison with the experiment. These data are
provided by {\em ab initio} numerical calculations using the
coupled channel scheme described in \cite{Servaas04}. The result
of this calculation for the elastic scattering cross-section is
presented in Fig. \ref{FigClosedChannel}. The most striking
feature of this figure is that it displays two peaks instead of
one, as predicted by Eqn. (\ref{EqnCrossSection}). This difference
can be easily understood by noting that the dipolar interaction
that couples the molecular state to the outgoing channel provides
a ``spin-orbit" coupling that modifies the relative orbital
angular momentum of the pair \cite{Ticknor04}. In other word, each
resonance corresponds to a different value of the relative angular
momentum $m_l$ (the $m_l=+1$ and $m_l=-1$ resonances are
superimposed because the frequency shift induced by the dipolar
coupling is proportional to $m_l^2$, as noted in
\cite{Ticknor04}).

As the spin-orbit coupling is not included in our simplified
three-level model, we take the multiple peak structure into
account by fitting the data of Fig. (\ref{FigClosedChannel}) using
a sum of three laws (\ref{EqnCrossSection}) with a different set
of phenomenological parameters for each value of the angular
momentum:

\begin{equation}
\sigma_{\rm
el}(E)=\sum_{m_l=-1}^{+1}\frac{C_{m_l}E^2}{(E-\tilde\delta_{m_l})^2+\hbar^2\gamma_{m_l}^2/4},
\label{EqnCrossSection2}
\end{equation}

\noindent where $\tilde\delta_{m_l}$ is related to the magnetic
field through $\tilde\delta_{m_l}=\mu (B-B_{F,m_l})$ and
$\gamma_{m_l}=\gamma_{0,m_l}+a_{m_l} E^{3/2}$. Using this law, we
obtain a perfect fit to the  elastic as well as inelastic data
obtained from the coupled channel calculations. The values
obtained for the different phenomenological parameters are
presented on Table \ref{table1}.

From these data, we see  first that the ``elastic" properties are
independent of $m_l$. This comes from the fact that the elastic
scattering is mainly a consequence of the hyperfine coupling that
does not act on the center of mass motion of the atoms. However,
we see that both the inelastic collision rate constant $D$ and the
molecule lifetime $\gamma_0$ exhibit large variations with the
relative angular momentum \cite{BiasNote}. First, the spontaneous
decay rate $\gamma_0$ of a molecule in $m_l=0,+1$ is always larger
than $\sim 10^{2}\,{\rm s}^{-1}$, which corresponds to a maximum
lifetime of about 10~ms. Second, we observe a strong reduction of
the losses in the $m_l=-1$ channel, in which no significant
spontaneous decay could be found. An estimate of $\gamma_0$ can
nevertheless be obtained by noting that, since the elastic
parameters are independent of $m_l$, the ratio $D/\gamma_0$ should
not depend on $m_l$ (this can be checked by comparing the ratios
$D/\gamma_0$ for $m_l=0$ and $m_l=+1$ in the (-1/2,-1/2) channel).
Using this assumption we find that $\gamma_0\sim 4\times
10^{-3}\,{\rm s}^{-1}$ both in (1/2,-1/2) and (-1/2,-1/2). The
reason for this strong increase of the lifetime of the molecules
in $m_l=-1$ is probably due to the fact that due to angular
momentum conservation
 the outgoing pair is expected to occupy a state with $l=3$
after an inelastic process. Indeed, if we start in a two-body
state $(m_F,m_F')$ and if the dipolar relaxation flips the spin
$m_F'$, then the atom pair ends up in a state $(m_F,m_F'+1)$. This
increase of the total spin of the pair must be compensated by a
decrease of the relative angular momentum. Therefore, if the
molecule was associated to a relative angular momentum $m_l$, it
should end up with $m_l-1$. In the case of  $m_l=-1$, this means
that the final value of the relative angular momentum is $m_l=-2$,
ie. $l\ge 2$. But, according to selection rules associated to
spin-spin coupling, the dipolar interaction can only change $l$ by
$0$ or $2$. Therefore, starting from a p-wave ($l=1$) compound,
this can only lead to  $l=3$.  Let us now   assume that the
coupling between the molecular state and the outgoing channel is
still proportional to the overlap between the two states (see Eqn.
\ref{EqnOverlap}), even in the presence of a dipolar coupling: the
argument above indicates that the ratio
$\rho=\gamma_{0,m_l=-1}/\gamma_{0,m'_l\not = -1}$ between the
decay rate of molecules in $m_l=-1$ and the one of molecules in
$m'_l\not =1$ is then of the order of

$$\rho\sim \left|\frac{\int g^* (r) j_3 (kr)r^2 dr}{\int g^* (r) j_1 (kr)r^2 dr}\right|^2,$$

\noindent where $k=\sqrt{m\Delta/\hbar^2}$ is the relative
momentum of the atoms after the decay.
For lithium, we have $R_e\sim 3$~\AA\, \cite{ReNote} which yields
$kR_e\sim 7 \times 10^{-2}$. This permits to approximate the
spherical Bessel function $j_l$ by their expansion at low $k$,
$j_l (kr)\sim (kr)^l$, that is

$$\rho\sim k^4 \left|\frac{\int g^* (r) r^5 dr}{\int g^* (r) r^3 dr}\right|^2 \sim \left(kR_e\right)^4,$$

With the numerical  value obtained for $kR_e$, we get $\rho\sim
2\times 10^{-5}$, which is, qualitatively, in agreement with the
numerical coupled channels calculations presented above.

\section{Temperature averaging}

In realistic conditions, the two body loss rate $G_2$ needs to be
averaged over the thermal distribution of atoms. $G_2$ is
therefore simply given by

$$G_2(E)=\sqrt{\frac{\pi}{4(k_{\rm B}T)^3}}\int_0^{\infty}{g_2 (E) e^{-E/k_{\rm B}T}E^{1/2}dE}.$$

\begin{figure}
\includegraphics[width=\columnwidth]{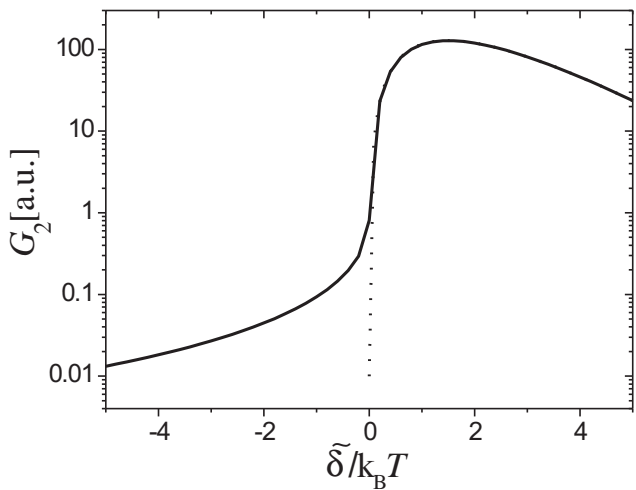} \caption{Full line: numerical calculation
of the loss-rate for $T=10~\mu$K. Dotted line: asymptotic
expansion (\ref{EqnG2p}).} \label{FigLoss}
\end{figure}

The evolution of $G_2$ vs detuning is displayed in Fig.
\ref{FigLoss} and shows a strongly asymmetric profile that was
already noticed in previous theoretical and experimental papers
\cite{Regal03,Ketterle}.

This feature can readily be explained by noting that in situations
relevant to experiments, $\gamma_0$ is small with respect to
temperature.  We can therefore replace $g_2$ by a sum of  Dirac
functions centered on $\delta_{0,m_l}$. When the $\delta_{0,m_l}$
are positive, $G_2$ takes the simplified form

$$G_2=4\sqrt{\pi}\sum_{m_l}\left(\frac{D_{m_l}}{\hbar\gamma_{m_l}(\tilde\delta_{m_l})}\right)\left(\frac{\tilde\delta_{m_l}}{k_{\rm
B}T}\right)^{3/2}e^{-\tilde\delta_{m_l}/k_{\rm B}T}.$$

Moreover, if we neglect the lift of degeneracy due to the dipolar
interaction coupling and  we assume all the $\tilde\delta_{m_l}$
to be equal to some $\tilde\delta$ , we get

\begin{equation}G_2=4\sqrt{\pi}\left(\bar\frac{D}{\hbar\bar\gamma
(\tilde\delta)}\right)\left(\frac{\tilde\delta}{k_{\rm
B}T}\right)^{3/2}e^{-\tilde\delta/k_{\rm B}T}, \label{EqnG2p}
\end{equation}

\noindent with $\bar D=\sum_{m_l} D_{m_l}$ and $\bar
D/\bar\gamma=\sum_{m_l}D_{m_l}/\gamma_{m_l}$ \cite{Convention}.

For $\tilde\delta<0$, and $|\tilde\delta|\gg k_{\rm B}T$, we can
replace the denominator of $g_2$ by $\tilde\delta^2$ and we get
the asymptotic form for $G_2$

\begin{equation}
G_2=\frac{3}{2}k_{B}T\sum_{m_l}\frac{D_{m_l}}{\tilde\delta_{m_l}^2}.
\label{EqnG2n}
\end{equation}

Let us now comment the two equations (\ref{EqnG2p}) and
(\ref{EqnG2n}).

\begin{enumerate}
\item We note that the maximum value of $G_2$ is obtained for
$\tilde\delta/k_{\rm B}T=3/2$. It means that when tuning the
magnetic field (ie., $\tilde\delta$), the maximum losses are not
obtained  at the resonance $\tilde\delta=0$, but at a  higher
field, corresponding to $\tilde\delta=3k_{\rm B}T/2$. For a
typical experimental temperature $T=10~\mu$K, this corresponds to
a shift of about $0.1$\,G.

\item Similarly, the width of $G_2 (\tilde\delta)$ scales like
$k_{\rm B}T$. Expressed in term of magnetic field, this
corresponds to  $\sim 0.1$\,G for $T=10\,\mu$K. This width is a
consequence of the resonance nature of the scattering
 in p-wave processes. As seen earlier, both elastic and inelastic collisions are more
 favorable when the relative energy $E=\tilde\delta$. When
 $\tilde\delta<0$, the resonance conditions cannot be fulfilled,
 since there are no state in the incoming channel  with negative
 energy. The scattering is then
formally analogous to optical pumping or other second-order
processes and yields the $1/\tilde\delta^2$ obtained in
(\ref{EqnG2n}). When $\tilde\delta\gg k_{\rm B}T$, the
 resonance condition is fulfilled by states that are not populated
 (since for a thermal distribution, we populate states up to $E\sim k_{\rm
 B}T$).

\end{enumerate}

\section{Conclusion}

In this paper, we have developed a  simple model capturing the
main scattering properties close to a p-wave Feshbach resonance.
The analytical formulas we obtained show very good agreement with
both numerical coupled channel calculations and experimental
measurements from our group \cite{Zhang04} and from the Innsbruck
group \cite{Chin04}. We have shown that the line shape of the
resonance is very different from what is expected for a s-wave
process: while s-wave scattering is mainly dominated by low energy
processes, p-wave scattering is rather dominated by collisions at
energies equal to that of the molecular state. Regarding p-wave
molecules, we have seen that at resonance their wave-function  was
dominated by the short range bare molecular state. Finally, the
study of the spontaneous decay of these molecules has shown a very
different lifetime depending on the relative angular momentum of
its constituents, since molecules in $m_l=-1$ could live $10^4$
times longer than in $m_l=0,+1$. This very intriguing result might
prove to be a valuable asset for the experimental study of p-wave
molecule since it guaranties that $m_l=-1$ dimers are very stable
against two-body losses in the absence of depolarizing collisions.

We thank Z. Hadzibabic, J. Dalibard and Y. Castin for very helpful
discussions. S.K.~acknowledges support from the Netherlands
Organisation for Scientific Research (NWO) and E.U. contract
HPMF-CT-2002-02019. E.K.~acknowledges support from the Stichting
FOM, which is financially supported by NWO. M. Teichmann
acknowledges support from  E.U. HPMT-2000-00102 and
MEST-CT-2004-503847 contracts. This work was supported by CNRS, by
the A.C.I. photonique and nanosciences programs from the french
ministry of research, and by Coll\`ege de France. Laboratoire
Kastler Brossel is {\it Unit\'e de recherche de l'\'Ecole Normale
Sup\'erieure et de l'Universit\'e Pierre et Marie Curie,
associ\'ee au CNRS.}


\begin{thebibliography}{90}
\bibitem{Convention} For the sake of simplicity, these  averaged parameters only were presented in
\cite{Zhang04}.
\bibitem{BiasNote} Note that the different lifetimes calculated for
the three $m_l$ states do not violate any rotational symmetry.
Indeed, the magnetic field used to reach the Feshbach resonance
provides a preferred direction to the system.
\bibitem{ReNote} We evaluate  $R_e$  by identifying it with the relative distance $r$
at which the total (centrifugal + long range) potential
$3\hbar^2/mr^2-C_6/r^6-C_8/r^8-C_{10}/r^{10}$ cancels and we used
the $C_{8,9,10}$ coefficients of \cite{Yan94}.
\end{thebibliography}
\end{document}